\documentclass[fullpage]{article}
\pdfoutput=1 
\usepackage[affil-it]{authblk} 
\usepackage{etoolbox}
\usepackage{lmodern}
\usepackage[a4paper]{geometry}
\usepackage[utf8]{inputenc}
\usepackage{amsmath}
\usepackage{graphicx}
\usepackage{lipsum}
\usepackage[dvipsnames]{xcolor}
\usepackage[colorlinks]{hyperref}
\hypersetup{linkcolor=blue,citecolor=blue,urlcolor=blue}

\newcommand{\Odd}{O(d,d)}

\newcommand{\ap}{\alpha'}

\makeatletter
\patchcmd{\@maketitle}{\LARGE \@title}{\fontsize{16}{19.2}\selectfont\@title}{}{}
\makeatother

\title{$\ap$-Cosmology: solutions and stability analysis}

\author[1]{Heliudson Bernardo\footnote{Email: \href{mailto:heliudson@hep.physics.mcgill.ca}{heliudson@hep.physics.mcgill.ca}}}
\author[2,3]{Guilherme Franzmann\footnote{Email: \href{mailto:guilherme.franzmann@su.se}{guilherme.franzmann@su.se}}}

\affil[1]{Physics Department, McGill University, Montreal, QC, H3A 2T8, Canada}
\affil[2]{The Oskar Klein Centre, Department of Physics, Stockholm University, AlbaNova, SE-10691 Stockholm, Sweden}
\affil[3]{Nordita and KTH Royal Institute of Technology, Roslagstullsbacken 23, SE-106 91 Stockholm, Sweden}

\date{}

\begin{document}

\maketitle

\begin{abstract}

We review O$(d,d)$ Covariant String Cosmology to all orders in $\ap$ in the presence of matter and study its solutions. We show that the perturbative analysis for a constant dilaton in the absence of a dilatonic charge does not lead to a time-independet equation of state. Meanwhile, the non-perturbative equations of motion allow de Sitter solutions in the String frame parametrized by the equation of state and the dilatonic charge. Among this set of solutions, we show that a cosmological constant equation of state implies a de Sitter solution both in String and Einstein frames while a winding equation of state implies a de Sitter solution in the former and a static phase in the latter. We also consider the stability of these solutions under homogeneous linear perturbations and show that they are not unstable, therefore defining viable cosmological scenarios.

\end{abstract}


\section{Introduction}

The very early universe remains the most conceivable laboratory for a theory of Quantum Gravity (QG). Assuming a smooth geometric phase, its dynamics can be divided into a background cosmology and perturbations on top of it. Typically, it is the latter that is encoded in observations like the Cosmic Microwave Background (CMB), which is a map of the fluctuations of the matter content that were present at the time of recombination. Given current observational windows, what the CMB tells us about the very early universe is encoded into two parameters: the amplitude of these fluctuations, $\mathcal{A}$, and the deviation from scale invariance in their power spectrum, $\mathcal{P}_S$, parametrized by the spectral index $n_s$. Neither of them provides us with a direct access to the underlying background dynamics that was evolving these fluctuations before the CMB was emitted \cite{Sunyaev:1970eu,Peebles:1970ag,Brandenberger:2019jbs}. 

Moreover, the fluctuations imprinted in the CMB are deep in the infrared in relation to QG's scale. That means that their dynamics can be treated approximately by the linear perturbation theory in General Relativity (GR). Within this paradigm, the $\mathcal{P}_S$ of these fluctuations is fully determined by the cosmological background evolution, parametrized by the Hubble parameter $H(t)$, and their initial conditions (IC). QG may play a direct role into both of them, for instance by deforming \cite{Martin:2000xs,Martin:2000bv} or explaining the IC (e.g. \cite{Halliwell:1984eu}) or by altering the background dynamics, i.e., modifying Friedmann equations (\cite{Ashtekar:2011ni,Gasperini:2002bn,Gasperini:2007zz} and references therein). 

Within the space of models to account for the observed values of $\{\mathcal{A},n_s\}$, inflationary cosmology remains the most successful proposal \cite{Guth:1980zm,Starobinsky:1980te,Sato:1980yn,Linde:1981mu,Albrecht:1982wi}. It considers the IC to be quantum mechanical while its background evolution is quasi-de Sitter, $\dot{H}(t) \approx 0$, which suffices to provide the necessary ingredients for an almost scale-invariant $\mathcal{P}_S$. In the scope of inflation, QG may change the IC for ultraviolet (UV) modes \cite{Brandenberger:2012aj} or even constrain how much it lasts \cite{Bedroya:2019tba}. All in all, inflation has largely motivated to seek for de Sitter (dS) like solutions in UV completions of gravity. 

Among the different QG coexisting approaches, String Theory remains the most successful proposal to this date. As expected, most of the early universe physics considered in its framework have been dS oriented \cite{Kachru:2003aw,Kachru:2003sx,Baumann:2014nda}.  However, alternative models can also be constructed. String Gas Cosmology (SGC) \cite{Brandenberger:1988aj}, for instance, is a toy model that considers the fluctuations to be thermal in their origin, and their stringy nature guarantees them to have a holographic scaling \cite{Brandenberger:2008nx,Battefeld:2005av}, while the background evolution is almost quasi-static, $H(t) \approx 0$. In more generality, SGC fulfills the requisites demanded by the emergent scenario discussed in \cite{Brandenberger:2019jbs}, which also provides the necessary elements in order to recover an almost scale-invariant $\mathcal{P}_S$. 

Given that the fluctuations' dynamics can be accounted by the linear perturbation theory in GR, our interest is to develop further the background cosmology in the context of String Theory. Our approach consists of looking to its universal NS-NS massless sector in a cosmological background including all classical string corrections, namely $\ap$-corrections, that are fixed by an $O(d,d)$ symmetry which is intrinsic to such a background. 

Historically, this particular avenue of research goes back to the early 90's after the realization that the cosmological background provided by these fields had a scale-factor duality \cite{Veneziano:1991ek,Tseytlin:1991xk}, which was later generalized to a global $O(d,d)$ symmetry \cite{Meissner:1991zj, Meissner:1991ge}. These developments were considered in the context of $0^{\text{th}}$ order in $\ap$-corrections and  an explicitly $O(d,d)$ covariant formalism was developed, with a manifestly duality invariant action \cite{Gasperini:1991ak}. Then, it was showed that this symmetry remained even after considering the full tower of $\ap$-corrections \cite{Sen:1991zi}  as long as the fields remained spatially independent. Although the $O(d,d)$ transformations also received corrections, in \cite{Meissner:1996sa} it was shown that the form of the transformations were preserved, at least to $1^{\text{th}}$ order in $\ap$. Recently, assuming this to be true to all orders, the form of the $\alpha'$-corrections were found, allowing a suitable classification of these corrections \cite{Hohm:2019jgu}\footnote{Cosmological applications in the absence of matter were considered in \cite{Wang:2019kez,Wang:2019dcj}.}. A natural extension of their results involved the inclusion of a matter sector \cite{Bernardo:2019bkz}, which finally allows us to consider realistic cosmological solutions.

Our goal in this paper is to briefly review $O(d,d)$ String Cosmology including $\ap$-corrections \cite{Hohm:2019jgu,Bernardo:2019bkz}, then to present and discuss some of its cosmological solutions and finally to consider their stability. This will open precedent to consider realistic cosmological scenarios in future works and to investigate the feasibility of the models discussed above and beyond.

The paper is organized as follows. In Section \ref{sec:review} we review the formalism and write down the equations of motion. Then, in Section \ref{sec:solutions} we introduce cosmological solutions for which a stability analysis will be considered in Section \ref{sec:stability}. Finally, in Section \ref{sec:conclusions} we conclude. 

\section{$\Odd$ String Cosmology including $\ap$- corrections} \label{sec:review}

The massless NS-NS sector of all superstring theories includes the metric, $G_{\mu\nu}(x)$, the Kalb-Rammond 2-form field, $B_{\mu\nu}(x)$, and the dilaton field, $\phi(x)$. For a cosmological background, we consider the fields given as $G_{00} = -n^2(t)$, $G_{0i} = 0$, $G_{ij} = g_{ij}(t)$, $B_{00} = 0$, $B_{0i} = 0$, $B_{ij} = b_{ij}(t)$ and $\phi = \phi(t)$. Then, the action for these fields coupled to matter including $\ap$-corrections to all orders, which are constrained by O$(d,d)$ symmetry, has been shown to have the form \cite{Bernardo:2019bkz} 
\begin{align}
    S &= \frac{1}{2\kappa^2}\int d^d x dt n e^{-\Phi}\left[-(\mathcal{D}\Phi)^2+X(\mathcal{DS})\right] + S_{m}[\Phi, n, \mathcal{S}, \chi]
\end{align}
where $\mathcal{D} \equiv 1/n \partial_t$ is the covariant temporal derivative, $\Phi \equiv 2\phi - \ln \sqrt{\det g}$ is the shifted dilaton, $\chi$ represents the matter sector and $X(\mathcal{DS})$ depends only on the first derivative of the $2d\times2d$ matrix
\begin{equation}
    \mathcal{S} = \eta \mathcal{H} = \begin{pmatrix}
    bg^{-1} & g- bg^{-1}b \\
    g^{-1} & - g^{-1}b 
    \end{pmatrix},
    \end{equation}
where
\begin{equation}
    \eta  = \begin{pmatrix}
    0 & 1 \\
    1 & 0 
    \end{pmatrix}, \quad 
    \mathcal{H} = \begin{pmatrix} 
    g^{-1} & -g^{-1}b \\ 
    bg^{-1} & g - bg^{-1}b
    \end{pmatrix},
\end{equation}
with $\eta$ being an $\Odd$ metric and $\mathcal{H}\in\Odd$, such that $\mathcal{S}^2 = 1$, while $\Phi$ is a scalar under $\Odd$ transformations. Considering only single trace contributions to $X(\mathcal{DS})$, that was shown in \cite{Hohm:2019jgu} to be enough for applications in flat FRLW cosmologies, the action can be written as
\begin{equation}
    S= \frac{1}{2\kappa^2}\int d^d x dt n e^{-\Phi}\left[-(\mathcal{D}\Phi)^2 + \sum_{k=1}^{\infty}\alpha'^{k-1}c_k \text{tr}(\mathcal{DS})^{2k}\right] + S_{m}[\Phi, n, \mathcal{S}, \chi], \quad c_1 = -\frac{1}{8} \label{eq:The_Action_2}
\end{equation}
which is manifestly invariant under the global $\Odd$ transformations and time reparameterizations. The values of the coefficients $\{c_k\}$ for $k>1$ are not known in general and take different values for different string theories ($c_2 = 1/64$ for heterotic strings and it vanishes for type II strings, for instance).

The equations of motion (EOM) are given by 
\begin{subequations}\label{eq:EOM_matter}
    \begin{equation}\label{eq:EOM_Phi_matter}
         2\mathcal{D}^2{\Phi} - (\mathcal{D}\Phi)^2 - \sum_{k=1}^{\infty}\alpha'^{k-1}c_k \text{tr}(\mathcal{DS})^{2k} = \kappa^2 e^{\Phi} \bar{\sigma},
    \end{equation}
    \begin{equation}\label{eq:EOM_n_matter}
        (\mathcal{D}\Phi)^2 - \sum_{k=1}^{\infty}\alpha'^{k-1}(2k-1)c_k \text{tr}(\mathcal{DS})^{2k} = 2\kappa^2 \Bar{\rho}e^{\Phi},
    \end{equation}
    \begin{equation}\label{eq:EOM_S_matter}
         \mathcal{D}\left(e^{-\Phi}\sum_{k=1}^{\infty}\alpha'^{k-1}4kc_k\mathcal{S}(\mathcal{DS})^{2k-1}\right) = -\kappa^2 \eta \Bar{\mathcal{T}},
    \end{equation}
\end{subequations}
where we defined a dilatonic charge \cite{Gasperini:2007ar}
\begin{equation}
    \sigma \equiv -\frac{2}{\sqrt{-G}}\frac{\delta S_m}{\delta \Phi},
\end{equation}
the bar denotes multiplication by $\sqrt{g}$, $\bar{\sigma} \equiv \sqrt{g}\sigma$, and we defined the O$(d,d)$ covariant energy-momentum tensor 
\begin{equation}
    \mathcal{\Bar{T}} \equiv \frac{1}{n}\left(\eta\frac{\delta S_m}{\delta \mathcal{S}} \mathcal{S}- \eta \mathcal{S}\frac{\delta S_m}{\delta \mathcal{S}}\right),
\end{equation}
while the energy density is defined as usual, 
\begin{align}
    \rho = \frac{\bar{\rho}}{\sqrt{g}} = \frac{T_{00}}{n^2}, \quad 
    T_{\mu\nu} = -\frac{2}{\sqrt{-G}}\frac{\delta S_m}{\delta g^{\mu\nu}} .
\end{align} 
From the EOM, we can also write down a continuity equation,
\begin{equation}
    \mathcal{D}{\Bar{\rho}} + \frac{1}{4}\text{tr}(\mathcal{S (D S)}\eta \mathcal{\bar{T}}) - \frac{1}{2}\bar{\sigma}\mathcal{D}{\Phi} = 0. \label{eq:covariant_general_continuity}
\end{equation}

Restricting ourselves to a flat FRLW background, meaning we consider a vanishing two-form, $n(t)=1$ and $g=a^2(t)I$, and taking the matter sector to be given by a perfect fluid defined as $T_{00}=\rho$, $T_{0i}=0$ and $T_{ij}=pg_{ij}$, the equations reduce to the $\ap$-corrected Friedmann equations
\begin{subequations}\label{eq:Alpha_Fried_eqs}
    \begin{equation}
    \dot{\Phi}^2 + HF'(H) - F(H) = 2\kappa^2 e^{\Phi}\Bar{\rho}   \label{eq:Alpha_Fried_eqs_a}
    \end{equation}
    \begin{equation}
     \dot{H}F''(H) - \dot{\Phi}F'(H) = -  2d \kappa^2e^{\Phi} \bar{p}   \label{eq:Alpha_Fried_eqs_b}
    \end{equation}
    \begin{equation}
      2 \Ddot{\Phi} - \dot{\Phi}^2 + F(H) = \kappa^2 e^{\Phi} \bar{\sigma},   \label{eq:Alpha_Fried_eqs_c}
    \end{equation}
\end{subequations}
where $p$ is the pressure, $H(t)$ is the Hubble parameter, $'$ denotes derivatives w.r.t. $H$ and the function $F(H)$ is defined as 
\begin{equation}
    F(H) = 2d \sum_{k=1}^{\infty}(-\alpha')^{k-1}c_k 2^{2k} H^{2k}.
\end{equation}
As expected, these equations are invariant under the scale factor duality transformation $a\rightarrow 1/a$, since under this transformation we have
\begin{equation}
    H\rightarrow-H,\quad \Phi \rightarrow \Phi, \quad f(H) \rightarrow - f(H), \quad g(H) \rightarrow g(H), \quad \bar{\rho}\rightarrow \bar{\rho}, \quad \bar{p} \rightarrow - \bar{p}, \quad \bar{\sigma} \rightarrow \bar{\sigma},
\end{equation}
which relies on the fact that the matter action is duality invariant, which is the case if one considers it to be given by a gas of free strings \cite{Gasperini:1991ak, Tseytlin:1991xk}. This is a remnant of the O$(d,d)$ symmetry in a FLRW background.

The continuity equation (\ref{eq:covariant_general_continuity}) reduces to
\begin{equation}
     \frac{\dot{\bar{\rho}}}{\bar{\rho}} + d H w -\frac{\lambda}{2}\dot{\Phi} = 0, \label{eq:continuity_equation}
\end{equation}
where we have introduced a barotropic equation of state (EOS), $w\equiv p / \rho$, and the density ratio (DR), $\lambda \equiv \sigma / \rho$, which measures how strongly the matter is coupled to the dilaton in relation to the metric \cite{Angus:2019bqs}. Throughout this paper, we will assume $\{w,\lambda\}$ to be constants, thus we can solve the continuity equation for the energy in terms of the scale factor and shifted dilaton,
\begin{equation}
    \bar{\rho} = \bar{\rho}_0 \left(\frac{a_0}{a}\right)^{dw} e^{\frac{\lambda}{2}(\Phi - \Phi_0)},
\end{equation}
where $\bar{\rho}_0$, $a_0$ and $\Phi_0$ are constants.

In the next sections we will study closely and systematically cosmological solutions to these equations. Since they are so far only written in the String frame (S-frame), it is also useful to write the Hubble parameter in the Einstein frame (E-frame), $H_E$, in terms of the S-frame variables. In \cite{Hohm:2019jgu}, $H_E$ as a function of the cosmic time in the E-frame, $t_E$, was shown to be
\begin{equation}
    H_E (t_E) = - (a(t))^{\frac{d}{d-1}}e^\frac{\Phi}{d-1} \frac{1}{d-1} \left(\dot{\Phi} + H\right), \quad   dt = dt_E e^{\frac{2\phi}{d-1}}\label{eq:Einstein_Hubble_parameter}
\end{equation}
and its evolution is given by
\begin{equation}
    \frac{dH_E(t_E)}{dt_E} = -\frac{e^{\frac{4\phi}{d-1}}}{d-1}\left[\ddot{\Phi}+\dot{H} + \frac{1}{d-1}(dH+\dot{\Phi})(H+\dot{\Phi})\right]. \label{eq:Hubble_evol_E_frame}
\end{equation}

\section{Cosmological Solutions}\label{sec:solutions}

The solutions to the $\ap$-Cosmological equations (\ref{eq:Alpha_Fried_eqs}) started to be considered in \cite{Bernardo:2019bkz}. Before we consider their stability, we will briefly review them and also introduce some new solutions that provide appealing cosmological scenarios. 

\subsection{Perturbative solution for a constant dilaton} 
    
It is known that at the lowest order in $\alpha'$, i.e., considering $c_k = 0$ for $k>1$ in (\ref{eq:Alpha_Fried_eqs}), the solution for constant dilaton and vanishing dilatonic charge implies a radiation EOS, $w = 1/d$ \cite{Tseytlin:1991xk}. In \cite{Bernardo:2019bkz} it was shown that upon turning on the $\ap$-corrections, it is not possible to obtain a solution with a constant EOS after imposing $\dot{\phi}=0=\sigma$. However, a perturbative solution was found,
\begin{align}
    H(t) &= \frac{H_0}{t} + \alpha' \frac{H_1}{t^3} + \alpha'^2 \frac{H_2}{t^5} + \dots,\\
    w(t) &= \frac{1}{d}-32 d  c_2 w_2 \alpha' H^2 +128 d c_3 w_3 \alpha'^2 H^4 -512 d c_4 w_4 \alpha'^3 H^6 + \dots,
\end{align}
where the EOS is time dependent instead and the coefficients $w_i$ and $H_i$ are completely fixed by the $c_k$ and the spacetime dimensionality. Note that, indeed, the first term corresponds to the radiation solution expected from the $0^{\text{th}}$ order equations. The above solution is potentially relevant for late-time cosmology since a rolling dilaton can lead to violations of the weak equivalence principle \cite{Gasperini:2007ar}, which is very constrained (see for instance \cite{Touboul:2017grn}), and its running may lead to a breakdown of the classical regime, given the dilaton modulates the strength of the string coupling.
    
\subsection{Static solution in S-frame}

If the Hubble parameter is zero, then $F(0) = F'(0) = 0$ and $F''(0) = 16 c_1 d$. Thus, (\ref{eq:Alpha_Fried_eqs_c}) gives as solution for the shifted dilaton,
\begin{equation}
    \dot{\Phi}(t) = -\frac{2}{t\left(1+ \frac{\lambda}{2}\right)+C_1},\quad \Phi(t) = C_2 - \frac{2}{1+\frac{\lambda}{2}}\ln \left[t\left(1+ \frac{\lambda}{2}\right)+C_1\right], \label{eq:sdil_sol_static}
\end{equation}
where $C_i$ are constants and $\lambda \neq -2$. Then, it is easy to see from (\ref{eq:Alpha_Fried_eqs_b}) that the EOS vanishes. Thus, a static solution is compatible with a pressureless equation of state. Note that for $H=0$ all the $\ap$-corrections vanish since they enter into the equations in powers of $H$. Therefore, this is a solution to the lowest order equations that remains valid even after including all $\alpha'$-corrections despite its energy scale.

This solution in the E-frame gives rise to 
\begin{equation}
    H_E(t_E) = \frac{2}{\left(2+ (d-1)(1+\lambda/2)\right)(t_E - t_{E,1})}, \quad a_E(t_E) \propto t^{\frac{2}{\left(2+ (d-1)(1+\lambda/2)\right)}}.
\end{equation}
Note that for $\lambda = 0$ it corresponds to a radiation dominated solution.

\subsection{Non-Perturbative de Sitter (dS) solutions} \label{subsec:non_pert_dS_sol}

It has been shown \cite{Bernardo:2019bkz} that after fixing $H=H_0$ to be constant with $\sigma = 0$ and constant EOS, the shifted dilaton satisfies $\dot{\Phi} = d w H_0$. It was also shown that if the condition for a constant $w$ is relaxed, then there is a dynamical solution with varying $\Phi$ and $w$ that reduces to the former case in the asymptotic limits $t\rightarrow \pm \infty$, with a different $w$ in each limit. In fact, as discussed in the next section, dS solutions of the form $\dot{\Phi} = -\beta H_0$ with constant $\beta$ are fixed points of the dynamical equations. This is true even for a non-zero DR,  for which $\beta$ depends on $\lambda$ as well. As we see below, for $\beta<0$ we find the solutions to be unstable. 

Among this class of dS solutions in the S-frame we find some special cases for $\lambda=0$. The first one comes about after analyzing the dilaton's velocity, which satisfies
\begin{equation}\label{eq:dot_dilaon_for_dS}
\dot{\phi} = \frac{d H_0}{2}(1+w).
\end{equation}
Thus, for $w = -1$, corresponding to a cosmological constant EOS, we get a dS solution in both frames since the dilaton is constant. The second special case can be found by looking to (\ref{eq:Einstein_Hubble_parameter}) and noticing that a static solution in the E-frame can be obtained by considering a winding EOS, namely $w=-1/d$. Thus, we have a time dependent dilaton solution with Minkowski metric in the E-frame. This opens precedent for considering the quasi-static phase required by String Gas Cosmology \cite{Brandenberger:1988aj} for the background cosmology and more generally for implementing the emergent scenario advocated in \cite{Brandenberger:2019jbs} as an alternative to inflation in the context of string cosmology. Both special cases are only possible due to the entire tower of $\ap$-corrections, therefore characterizing non-perturbative solutions. 

\section{General Stability Analysis} \label{sec:stability}

The covariant String Cosmology equations of motion with $\ap$-corrections are coupled and nonlinear. To proceed with the stability analysis, we treat the equations as a dynamical system and consider linear perturbations around a given solution, in a similar fashion to what was done in \cite{Quintin:2018loc} to handle first order $\ap$-corrections. In order to develop some intuition first, let's look to the dynamical analysis of the Friedmann equations in GR.

\subsection{Dynamical Analysis of Friedmann Equations} \label{subsec:analysis_GR}

For $D=1+3$ in the absence of spatial curvature, the dynamical equation after taking into account the constraint equation is given by
\begin{equation}
    \dot{H} = - \frac{3}{2} (1+w) H^2,
\end{equation}
assuming a barotropic EOS. We see that a fixed point, namely $\dot{H}=0$, is given by $w=-1$, a dS solution as expected, or vacuum, $H=0$. 

In order to look at the stability of the dS solution, let us write the dynamical equation as $\dot{H} = C(H)$ and look at fluctuations around the fixed point solution, $H_0$. In fact, this results into
\begin{equation}
    \dot{\delta H} = C'(H_0) \delta H. \label{eq:GR_pert}
\end{equation}
It is easy to see that $C'(H_0) = 0$, so any homogeneous perturbation in the dS solution will shift the value of the Hubble parameter. That does not mean that the dS solution is unstable though, it just means that a perturbation in the energy density has happened, as one can easily check after looking to the continuity equation. Both the energy density and the Hubble parameter shift by a constant amount\footnote{Another insightful exercise is to consider a power-law solution, $H=n/t$. Then, solving (\ref{eq:GR_pert}) shows that a crude phantom EOS, $w<-1$, leads to instabilities.}.

\subsection{Dynamical Analysis of $\alpha'$-Cosmology} \label{subsec:dynamical_analysis}

Following the same prescription, we need first to rewrite (\ref{eq:Alpha_Fried_eqs}) as a system of first order ordinary differential equations (ODE). Given we are interested in a relatively short, single phase evolution dominated by a certain type of matter, we consider a constant EOS and a constant DR. Thus, (\ref{eq:Alpha_Fried_eqs}) reduce to
\begin{subequations}
    \begin{equation}
        \dot{H} \equiv C_1(H,y) =  \frac{1}{F''}\left[ y F' - dw(y^2 +HF'-F) \right] \label{eq:ODE_1}
    \end{equation}
    \begin{equation}
        \dot{y} \equiv C_2(H,y) = \frac{y^2}{2} - \frac{F}{2} + \frac{\lambda}{4} (y^2 + HF'-F), \label{eq:ODE_2}
    \end{equation}
\end{subequations}
where $\dot{\Phi} \equiv y$ and we have assumed $F''(H) \neq 0$, together with the constraint equation now written as
\begin{equation}
    y^2 + H F' - F = 2 \kappa^2 e^\Phi \bar{\rho}.
\end{equation}

\subsubsection{Fixed Points}

The fixed points, also called equilibria, $X$, given by solutions of the ODEs where $\dot{H} = \dot{y}=0$, are defined by $C_i (X) = 0$, which implies
\begin{subequations} \label{subeq:fixed_point}
    \begin{equation}
        y_0F'_0 - dw(y^2_0 - F_0) -dwH_0 F'_0 =0 
    \end{equation}
    \begin{equation}
        \left(1 + \frac{\lambda}{2}\right) (y^2_0 - F_0) + \frac{\lambda}{2}H_0 F'_0=0.
    \end{equation}
\end{subequations}
where $H_0$ and $y_0$ are constants and $F^{(n)}_0 \equiv F^{(n)}(H_0)$.
They can be combined into 
\begin{equation}\label{eq:condition_for_fixed_point}
    F'_0\left[H_0 - \frac{y_0}{dw}\left(1 + \frac{\lambda}{2}\right)\right] =0,
\end{equation}
where we assumed $w\neq0$ (the case of a vanishing EOS will be considered later). The fixed points corresponds to dS solutions in the S-frame with a constant shifted dilaton's velocity.

\subsubsection{Stability around equilibria}

In order to analyse stability around the fixed points, we need to linearize the system of ODEs and then look at the eigenvalues of the matrix defined by $A \equiv \partial_{\{H,y\}} C_i (X)$ at the equilibria. The matrix is given by 
\begin{equation} \label{eq:eigenvalue_matrix}
    A = \begin{pmatrix}
     y_0 -dH_0w & \frac{1}{F''_0}(F'_0 - 2dwy_0) \\
    -\frac{F'_0}{2} + \frac{\lambda}{4}H_0 F''_0 & y_0\left(1 + \frac{\lambda}{2}\right) \end{pmatrix}.
\end{equation}
%
%
%
The cases in which we are mostly interested in this paper are all of the form $y_0 = - \beta H_0$, as summarized in Section \ref{subsec:non_pert_dS_sol}. Then, the eigenvalues of $A$ are given by, 
\begin{equation}
    \alpha_{\pm} = -\frac{H_0\beta}{2} - \frac{H_0}{2} \left(\beta + \frac{\beta \lambda}{2} + dw\right) \pm \frac{1}{2}\sqrt{H^2_0 \left[ \frac{\lambda F'_0}{H_0} + \left(dw + \frac{\beta \lambda}{2} \right)^2 \right] -  \frac{2F'_0}{F''_0} (F'_0 + 2dw\beta H_0)}.
\end{equation}
Unstable modes are characterized by positive eigenvalues, while decaying ones are associated to negative eigenvalues. When we have vanishing eigenvalues further analysis is required. In particular, that was the case for the dS solution in GR studied above in Section \ref{subsec:analysis_GR}, where the perturbations only cause a shift on the background parameters and do not lead to exponential instabilities. We will also find that among our solutions.

\subsection{Stability of $\ap$-cosmological solutions}

We are finally ready to consider the stability of the solutions introduced in Section \ref{subsec:non_pert_dS_sol}. From Section \ref{subsec:dynamical_analysis}, it is clear that they are fixed points solutions with $\dot{\Phi}_0 = -\beta H_0 = \text{constant}$, where $\beta = 1$ gives the static phase in the E-frame and $\beta=d$ gives the dS solution in both frames. The equations of motion imply (for $w\neq 0$)
\begin{align}
    \frac{H_0F'_{0}}{dw}\left[dw + \beta\left(1+\frac{\lambda}{2}\right)\right]=0\\
    \beta^2H_0^2 - F_{0} = 2\kappa^2 e^{\Phi_0} \bar{\rho_0}\left(1+\frac{dw}{\beta}\right),
\end{align}
where we have used the fact that for constant $\dot{\Phi}$, $w$ and $\lambda$  we have
\begin{equation}
    e^{\Phi(t)}\bar{\rho}(t) = e^{\Phi_0}\bar{\rho}_0.
\end{equation}
The condition for fixed points (\ref{eq:condition_for_fixed_point}) can now be written as
\begin{equation}
    dw + \beta \left(1 + \frac{\lambda}{2} \right) = 0, \label{eq:cond_for_fixed_point_2}
\end{equation}
where we have assumed $F'_0\neq0$\footnote{The solution with $F'_0= 0$ was considered in \cite{Bernardo:2019bkz}.}. The values of $F'_{0}$, $F_{0}$ and $H_0$ can be fixed by the EOM (for $1-\lambda/2\neq0$ and $\lambda \neq -1$)\footnote{In order to recover these relations, one can follow the prescription already outlined in \cite{Bernardo:2019bkz} in Section 6.2., where the case $\lambda=0$ was considered.}
\begin{align}
    F'_{0} &= \frac{2d^2w^2H_0}{1+\frac{\lambda}{2}}, \label{eq:F_p_dS}\\
    F_{0} &= \frac{1+\lambda}{2+\lambda} H_0 F'_{0}, \label{eq:F_dS}\\
    H_0^2 &= -\frac{1}{1+\frac{\lambda}{2}} \frac{\kappa^2e^{\Phi_0}\bar{\rho}_0}{dw\beta}.
\end{align}
From the last equation, we see that either $w>0$ and $\lambda < -2$ or $w<0$ and $\lambda > -2$. Our main focus is on the latter case. Note that the last equation shows that the Hubble parameter's scale is defined by the matter content, which does not have to be necessarily close to the string scale, potentially allowing for realistic phenomenological values.

To check the solutions' stability at fixed points, let us look at the sign of the eigenvalues at equilibria after considering (\ref{eq:cond_for_fixed_point_2})-(\ref{eq:F_dS}),  
%
%
%
\begin{equation}
    \alpha_\pm =  -\beta \frac{H_0}{2} \left(1 \mp \left| 1 + \frac{2dw}{\beta} \right| \right).
\end{equation}
For $\lambda=0$, then $\beta+dw=0$ and the eigenvalues are $\{0,-\beta H_0\}$: one mode is decaying and one is constant. The constant mode is not worrying, since it corresponds to exactly what happens when one considers perturbations around dS solutions in GR, now with a constant shift not only on the Hubble parameter but also in the shifted dilaton's velocity.

\begin{figure}[h!]
    \centering
    \includegraphics[scale=.48]{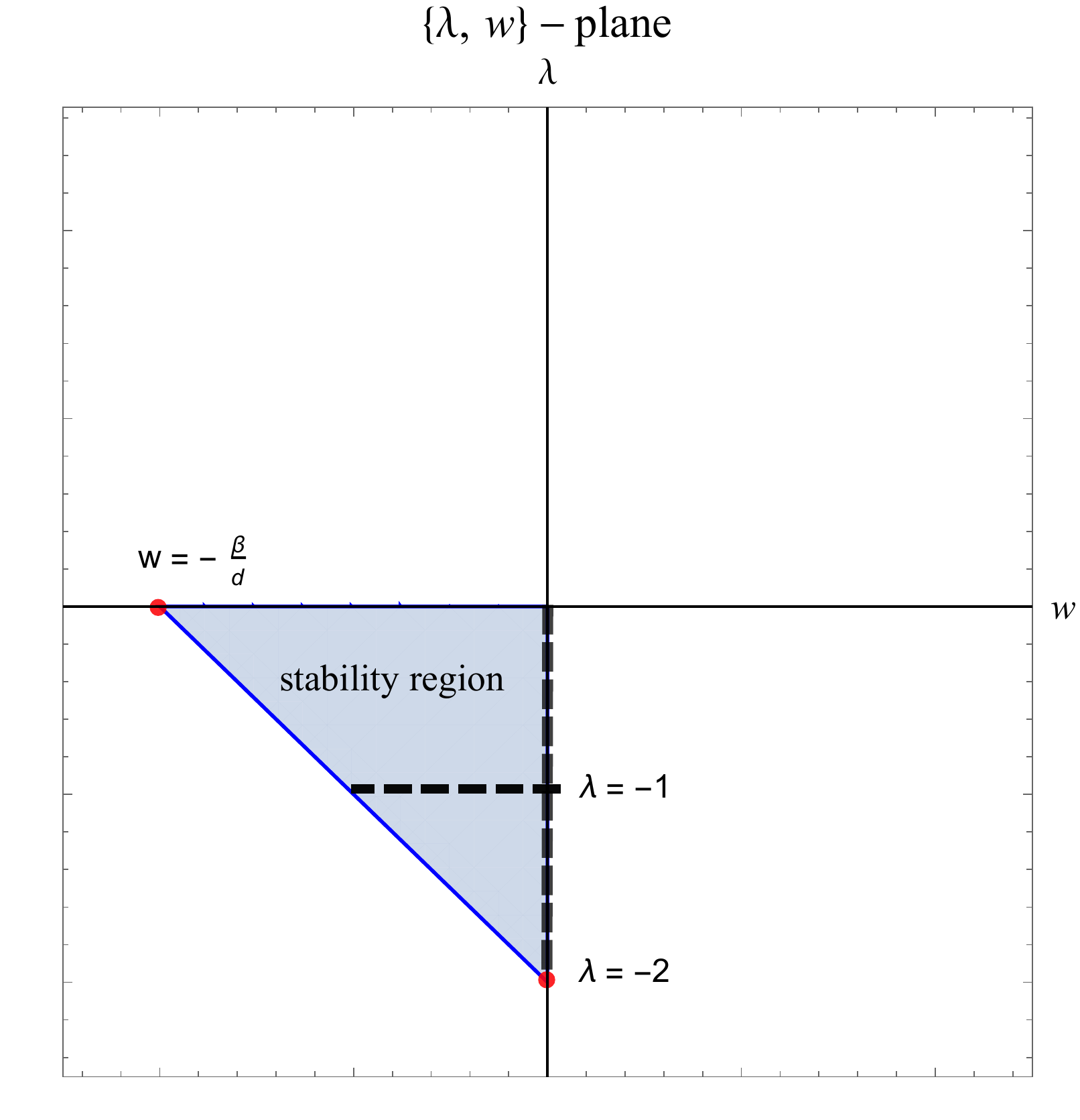}
    \caption{The shaded region indicates where the solutions are stable despite the properties of the function $F$. For $w=0$ or $\lambda=-1$, in general the stability is contingent on the sign of $F''(H_0)$, represented by the dashed line on these boundaries, while the case $\{w,\lambda\} = \{0,-2\}$ is stable as long as  $\dot{\Phi}<0$ and it can be grouped together with the class $\dot{\Phi}=-\beta H$ for $\beta>0$.}
    \label{fig:estability_plane}
\end{figure}

More generally, what we need to guarantee is that the eigenvalues are never positive. Thus, $\beta$ has to be positive and the term in the square root implies
\begin{equation}
    w\left(1 + \frac{dw}{\beta} \right) \leq 0,
\end{equation}
which implies $w \in [-\beta/d,0)$. Note that in the above analysis we have left out a few cases that need to be considered separately: $w=0$, $\lambda = \{ -1,-2 \}$. In order to check them, one can make direct use of (\ref{subeq:fixed_point}) and (\ref{eq:eigenvalue_matrix}). For $w=0$ and $\lambda \neq -2$, which also implies $y_0 = 0$ (since we are not interested in vacuum solutions), we find 
\begin{align}
    & H_0 F'_{0} = 2 \kappa^2 e^{\Phi_0} \bar{\rho}_0 \left(1+ \frac{\lambda}{2} \right), \quad  F_{0} = \lambda \kappa^2 e^{\Phi_0} \bar{\rho}_0\\
   &\alpha_{\pm} = \pm \frac{1}{2} \sqrt{- \frac{2 {F'_{0}}^{2}}{F''_0} + \lambda H_{0} F'_{0} },
\end{align}
so that the eigenvalues have opposite signs and this fixed point will be a saddle point as long as the term in the square roots is positive, otherwise the two eigenvalues will be imaginary and the perturbations are periodic, having a fixed amplitude (the fixed point is a neutrally stable center \cite{strogatz:2000}). While for $w=0$ and $\lambda=-2$ we find that the solution is stable as long as $y_0 = \dot{\Phi}_0<0$. For the remaining cases, the stability of the solutions is contingent on $F''_{0}$. Our findings are summarized in Figure \ref{fig:estability_plane}.

\subsection{Dynamical dS solutions}

The dS solutions with constant $\dot{\Phi}$ can be recovered from the asymptotic limit of the general solution with evolving $\dot{\Phi}$ and EOS. This was already shown in \cite{Bernardo:2019bkz} for the case $\lambda = 0$. For constant $\lambda > -2$, we may combine equations (\ref{eq:Alpha_Fried_eqs_a}) and (\ref{eq:Alpha_Fried_eqs_c}) to write the following equation for $\Phi(t)$
\begin{equation}
   \ddot{\Phi} - \frac{1}{2}\left(1+ \frac{\lambda}{2}\right)\dot{\Phi}^2 + \frac{1}{2}\left(1+ \frac{\lambda}{2}\right)F_{0} - \frac{\lambda}{2}H_0F'_{0} = 0,
\end{equation}
which is solved by
\begin{equation}
   \Phi(t) = \Phi_0 - \frac{4}{2+\lambda}\ln\left[\cosh\left(\frac{1}{4}\sqrt{(2+\lambda)|F_{0}(2+\lambda)- \lambda H_0 F'_{0}|}(t-t_0)\right)\right].
\end{equation}
Taking the asymptotic limits
\begin{equation}
    \lim_{t\to \pm \infty}\dot{\Phi} = \frac{\sqrt{|F'_0 H_0 \lambda -F_0 (\lambda +2)|}}{\sqrt{\lambda +2}} = \pm \frac{dw}{1+\frac{\lambda}{2}}H_0,
\end{equation}
where we used equations (\ref{eq:F_p_dS}) and (\ref{eq:F_dS}) to write the last equality for $w<0$. We see that we recovered the condition (\ref{eq:cond_for_fixed_point_2}) for a fixed point solution, also serving as a consistency check of our analysis. For $t\to \infty$, $\beta>0 $ and the asymptotic solution is a stable fixed point while for $t\to -\infty$ we have $\beta<0$ and the solution is unstable.

\section{Summary and Discussions} \label{sec:conclusions}

We have reviewed the action and equations of motion encompassing O$(d,d)$ Covariant String Cosmology to all orders in $\ap$. Since the formalism has been introduced in the String frame, we have also written down the evolution of the Hubble parameter in the Einstein frame so that typical cosmological scenarios could be considered and discussed. In order to check the viability of such scenarios, we have also gone through a linear stability analysis. The summary of the results obtained is as follows. 

The perturbative solution for a constant dilaton in the absence of a dilatonic charge implies a time-dependent equation of state for barotropic matter. Even though perturbatively the EOS is still completely fixed and it reproduces a radiation EOS at the $0^\text{th}$ order as it was expected from the lowest order EOM, higher $\ap$-corrections imply that a constant EOS is not a solution.

We have considered de Sitter solutions in the String frame and showed that they lead to a constant velocity for the shifted dilaton asymptotically, which depends on the asymptotic value of the EOS; in more generality, it also depends on the density ratio $\lambda$. Among the different values the EOS can take, $w=-1$ and $w=-1/d$ stand out when $\lambda=0$.

For a cosmological constant EOS, we have seen that the dilaton is constant, which contrasts with having a radiation EOS that was otherwise expected from the lowest order equations as discussed above. This solution implies an equivalence between String and Einstein frames, therefore implying a de Sitter solution also in the latter. Moreover, we have shown that the stability of this solution is on the same grounds as the typical de Sitter solution encountered in GR. 

Our most promising result occurs for a winding equation of state, $w=-1/d$. We have seen that there is a de Sitter solution in the String frame with this EOS that implies a static solution in the Einstein frame. Since first proposed, String Gas Cosmology relied on a quasi-static phase in the Einstein frame in order to recover an almost scale-invariant power spectrum, despite the fact that such a phase could not be realized through the evolution of the stringy EOS associated with a gas of closed strings after using the lowest order EOM in $\ap$. On the other hand, in $\ap$-Cosmology this phase is naturally realized by a winding EOS and that corresponds precisely to the initial phase of the stringy EOS. Given that this solution is also shown to not be unstable under a linear stability analysis, this can be seen as a proof of principle that the dynamics proposed in SGC, and more generally advocated in the emergent scenario, may be realized in String Cosmology after considering all $\ap$-corrections \cite{Bernardo:2019xx}. 

\section{Acknowledgements}

The authors thank Jerome Quintin and Robert Brandenberger for reading the manuscript and relevant discussions. H. B. is thankful to Nordita for the hospitality while this work was developed. The
research at McGill is supported by funds from NSERC.

\thispagestyle{plain}
\bibliographystyle{JHEP}
\bibliography{references}

\end{document}